\documentclass[structabstract]{aa}

\newcommand{\saxj}{SAX~J1808.4--3658\xspace}

\newcommand{\igrj}{IGR~J00291+5934\xspace}
\newcommand{\newigrj}{IGR~J17511--3057\xspace}

\newcommand{\swiftja}{Swift~J1756.9-2508\xspace}

\newcommand{\xtejd}{XTE~J1751--305\xspace}

\newcommand{\ngcj}{NGC 6440 X--2\xspace}

\newcommand{\rxte}{\textit{RXTE}\xspace}
\newcommand{\integral}{\textit{INTEGRAL}\xspace}
\newcommand{\swift}{\textit{Swift}\xspace}
\newcommand{\chandra}{\textit{Chandra}\xspace}

\newcommand{\fermi}{\textit{Fermi}\xspace}

\newcommand{\np}[2]{\ensuremath{{#1}\times10^{#2}}}

\newcommand{\Mdot}{\ensuremath{\dot{\rm M}}\xspace}

\newcommand{\Tstar}{\ensuremath{{\rm T}^{\star}}\xspace}
\newcommand{\Porb}{\ensuremath{{\rm P}_{\rm orb}}\xspace}
\newcommand{\Porbt}{\ensuremath{{\rm \dot{P}}_{\rm orb}}\xspace}
\newcommand{\ecc}{\ensuremath{\it e}\xspace}
\newcommand{\Msun}{\ensuremath{{\rm M}_\odot}\xspace}
\newcommand{\pap}{\citetalias{papitto_08}\xspace}
\newcommand{\markw}{\citetalias{markwardt_02}\xspace}

\newcommand{\changes}[1]{#1}

\usepackage{graphicx}
\usepackage{amsmath} 

\usepackage{txfonts}

\usepackage{subfigure}
\usepackage{xspace}
\usepackage{natbib}
\usepackage[english]{babel}
\usepackage{url}

\bibpunct{(}{)}{;}{a}{}{,}

\begin{document}

\title{Secular spin-down of the AMP \xtejd}

\author{A. Riggio\inst{1,2}, L. Burderi\inst{2}, T. {Di
    Salvo}\inst{3}, A. Papitto\inst{1,2}, A. {D'A\`\i}\inst{3},
  R. Iaria\inst{3} \and M. T. Menna\inst{4}}

\institute{INAF/Osservatorio Astronomico di Cagliari, localit\`a
  Poggio dei Pini, strada 54, 09012 Capoterra, Italy;
  \email{ariggio@oa-cagliari.inaf.it} \and Universit\`a~di Cagliari,
  Dipartimento di Fisica, SP Monserrato-Sestu km 0,7, 09042 Monserrato
  (CA), Italy \and Dipartimento di Scienze Fisiche e Astronomiche,
  Universit\`a~di Palermo, Via Archirafi 36, Palermo, 90123 Italy \and
  Osservatorio Astronomico di Roma, Sede di Monteporzio Catone, Via
  Frascati 33, Roma, 00040 Italy}

\authorrunning{A. Riggio et al.}

\date{}
\abstract
{Of the 13 known accreting millisecond pulsars, only a few showed more
  than one outburst during the \rxte era. \xtejd showed, after the
  main outburst in 2002, other three dim outbursts. 
  We report on the timing analysis of the latest one, occurred on
  October 8, 2009 and serendipitously observed from its very beginning
  by \rxte.}
{The detection of the pulsation during more than one outburst permits
  to obtain a better \changes{constraint} of the orbital parameters
  and their evolution as well as to track the secular spin frequency
  evolution of the source.}
{Using the \rxte data of the last outburst of the AMP \xtejd, we
  performed a timing analysis to improve the orbital parameters.
  Because of the low statistics, we used an epoch folding search
  technique on the whole data set to improve the local estimate of the
  \changes{time of} ascending node passage.}
{Using this new orbital solution we epoch folded data obtaining three
  pulse phase delays on a time span of 1.2 days, that we fitted using
  a constant spin frequency model.  Comparing this barycentric spin
  frequency with that of the 2002 outburst, we obtained a secular spin
  frequency derivative of \np{-0.55(12)}{-14} Hz s$^{-1}$.  In the
  hypothesis that the secular spin-down is due to a rotating magneto
  dipole emission, consistently with what is assumed for radio
  pulsars, we estimate the pulsar's magnetic dipole value.  We derive
  an estimate of the magnetic field strength at the polar cap of
  B$_{PC}$ = \np{4.0(4)}{8} Gauss, for a neutron star mass of
  1.4\Msun, \changes{assuming the Friedman Pandharipande Skyrme}
  equation of state.}
{}


\maketitle

\defcitealias{markwardt_02}{M02}
\defcitealias{papitto_08}{P08}

\section{Introduction}
\xtejd is one of the accretion powered millisecond X-ray pulsars
(AMPs) that showed more than one outburst in the \rxte era.
Recurrent outbursts were also observed in \saxj
\citep[][]{disalvo_08,burderi_09,hartman_09}, \igrj
(\citealt{galloway_05, galloway_atel_08, Patruno_10, Hartman_11,
  papitto_10}) and recently in \ngcj \citep[][]{altamirano_atel_10}
and \swiftja \citep[][]{Patruno_10b}.
\xtejd was detected for the first time by \rxte on April 3, 2002
(\citealt{markwardt_02}, \citetalias{markwardt_02} henceforth).
This outburst was the brightest and longest of the four showed by
\xtejd, permitting \citetalias{markwardt_02} and \citet{papitto_08} to
obtain a full orbital solution.

The second outburst, was spotted by \integral
\citep[][]{grebenev_atel_05} on March 28, 2005 and lasted $\sim$2 days
with a peak flux which was $\sim$ 14\% of the one reached during the
first outburst.
Unfortunately the \rxte PCA instrument was not in the proper mode to
detect pulsations during the first observation
\citep[][]{swank_atel_05}.
The follow-up observations was done in event and single bit modes, but
the source already fainted below detection.
The attribution of this outburst to \xtejd is not certain, since
pulsations were not detected and the \integral IBIS/ISGRI source
position is compatible with the position of at least \changes{one
  other} source, \newigrj\citep[][]{papitto_10, riggio_11}.

The third outburst was detected on April 5, 2007 by \rxte with a peak
flux of 18\% of the first outburst \citep[][]{markwardt_atel_07,
  falanga_atel_07}, very similar to the second outburst. 
In the subsequent pointed observation by \rxte the source became too
dim to detect pulsations. 
In this case the source identification is certain, thanks to a
simultaneous \swift observation \citep{markwardt_atel_07b}.

The latest outburst was first spotted with \integral
\citep{chevenez_atel_09}.
Fortunately, it occurred while \rxte was observing the last discovered
AMP \newigrj, very near to \xtejd \citep[][]{markwardt_atel_09b}.
The detection of the 435 Hz X-ray pulsation and a following \swift
observation confirmed that the source in outburst was \xtejd and not a
re-brightening of \newigrj.
In this paper we report on the timing analysis of this outburst.

\section{Observation and Data Analysis}
In this work we analyse \rxte PCA observations of \xtejd.
We use data from the PCA (proportional counter array, see
\citealt{Jahoda_06}) instrument on board of the \rxte satellite (ObsId
94041 and 94042).
We used data collected in event packing mode, with time and energy
resolution of $122 \mu s$ and 64 energy channels respectively.
\begin{figure}
  \includegraphics[width=\columnwidth]{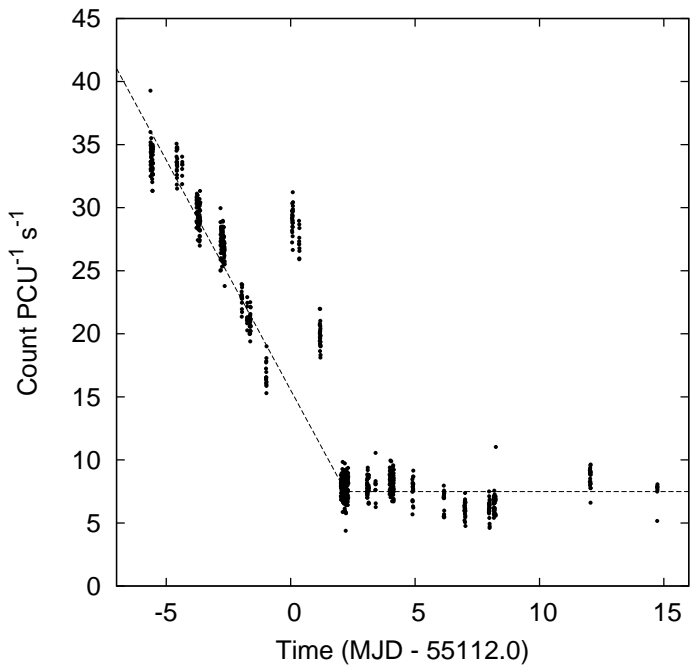}
  \caption{PCU 2 count rate (2-18 keV), subtracted of its background,
    is reported as a function of time in the period from 2 October
    2009 and 22 October 2009. During the first days, the last phase of
    the \newigrj outburst is visible. The flux re-brightening is
    caused by the onset of the \xtejd outburst, which lasted less than
    two days. The remaining days show the constant flux due to the
    galactic ridge. The superimposed model is the best-fit using a
    piecewise linear function. Since we are interested in determine
    the background due to \newigrj and the galactic ridge, we excluded
    from the fit the \xtejd outburst. See the text for more details.}
  \label{fig:flux}
\end{figure}
Although the \xtejd pulsation was detected only during observations
performed on 8 and 9 October 2009 \citep{markwardt_atel_09b}, the data
analysed here cover the time span from 6 October 2009 to 22 October
2009, as shown in Figure \ref{fig:flux}. This allows to precisely
determine any contribution to the observed emission attributable to
the AMP \newigrj, that was still in outburst during the \rxte
observation, and to the galactic ridge.
We obtained the energy band which maximises the signal to noise ratio
comparing the source X-ray spectrum and the background. We chose the
2--18 keV energy band.
We corrected the photon arrival times for the motion of the
earth-spacecraft system with respect to the solar system barycentre
and reported them to barycentric dynamical times at the solar system
barycentre using the \textit{faxbary} tool (DE-405 solar system
ephemeris), adopting the \chandra source position reported by
\citetalias{markwardt_02}. 
The uncertainty on the source position quoted by
\citetalias{markwardt_02} is 0.6$''$, $90 \%$ confidence level, 
as shown in Table \ref{table1}.

We obtain a first estimate of the mean spin frequency constructing a
Fourier power density spectrum of the 3.2 ksec exposure ObsID
94041-01-04-08 data calculating 53 power spectra from 64-s long data
segments ($2^{-11}$ bin size) which were averaged into one power
spectrum.
As reported in \citet{markwardt_atel_09b}, we found a signal at $\sim
435.32$ Hz.  No conclusive evidence of pulsations in the following
observations was found in this preliminary step.

\subsection{Determination of the local \Tstar}
A timing analysis to obtain an orbital solution at the time of the 
latest outburst from \xtejd was not possible because of the weakness 
of the pulsation.
However, it is still possible to correct the time series for delays 
induced by the orbital motion adopting the orbital parameters of the 
April 2002 outburst estimated by \citet{papitto_08} (\pap hereafter).
Propagating the uncertainty in the orbital period \Porb given by \pap
along the $\sim7$ years separating the 2009 outburst from the 2002
one, resulted in an uncertainty on the time of passage through the
ascending node, \Tstar, in the 2009 outburst of $\sim$ 186 s at
1$\sigma$ confidence level, about $\sim$ 7\% of the orbital period. 
Moreover the presence of an orbital period derivative \Porbt might
introduce a further shift on \Tstar. 
The true local value of \Tstar can thus be significantly different
from the nominal value obtained by propagating the orbital solution
provided by \pap.

To determine the best local orbital solution, we make the reasonable
hypothesis that the best local set of orbital parameters is the one
which gives the best signal to noise ratio, that is, in our case, the
highest $\chi^2$ value in an epoch folding search \citep[see
e.g.][]{kirsch_04}.
We restrict our search to just one orbital parameter, \Tstar, which is
the parameter with the largest uncertainty during the 2009
outburst. To explore all the possible values for \Tstar, the orbital
period being $\simeq$ 2546 s, we produced 2546 different time series
from the data of the 2009 outburst, which were corrected for the
orbital modulation. For each of these time series the adopted orbital
parameters were the same, except for \Tstar, which is varied in steps
of 1 s.  We then performed an epoch folding search for the spin period
on each of the 2546 time series using 32 phase bins to sample the
signal.
\begin{figure}
  \includegraphics[width=\columnwidth]{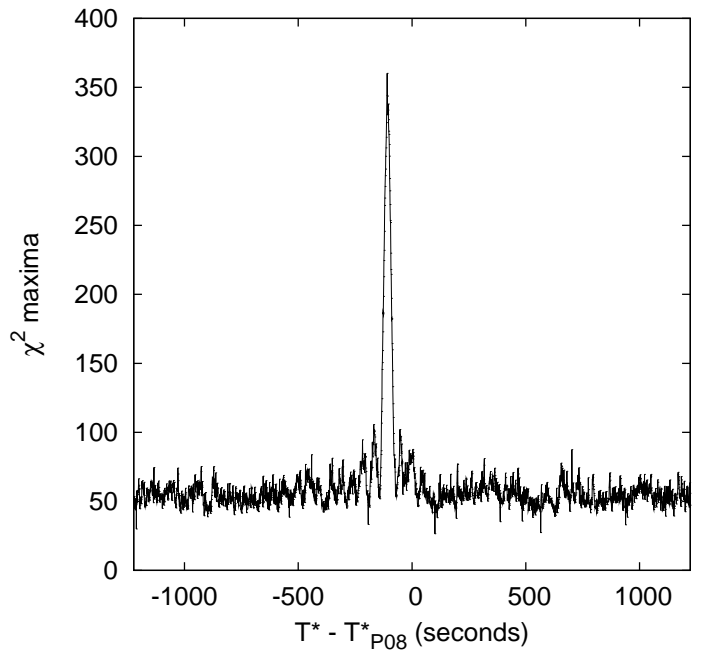}
  \caption{$\chi^2$ maxima obtained from an epoch folding search on
    the 2009 data corrected for the orbital modulation, varying each
    time the epoch of passage through the ascending node with 1 s
    step. A total of 2546 orbital solutions were tried, exploring all
    the possible values for \Tstar.}
  \label{fig:f1}
\end{figure}
In Fig. \ref{fig:f1} we show the maximum of the $\chi^2$ obtained from
the epoch folding search of each time series as a function of the
\Tstar adopted to produce the time series on which the folding search
has been performed.
We fitted the $\chi^2$ maxima curve with a model consisting of a
constant plus a Gaussian.
A clear peak at $\Delta\Tstar \simeq -110$ s is evident, well within
the $1\sigma$ confidence level of the orbital solution given by \pap
propagated to the 2009 outburst.
Thus we derived the time of passage from the ascending node during the
2009 outburst as $\Tstar_{09} = \Tstar_{P08} + \Delta\Tstar$.

Adopting the new value of \Tstar, we barycentered all the data
covering the 2009 outburst of \xtejd and performed again an epoch
folding search of the spin period. The improvement of the orbital
solution allowed us to detect the pulsation in two other observations,
corresponding to ObsID 94041-01-04-04 (MJD 55112.335, with an exposure
of 1.3 ksec), and ObsID 94042-01-02-00 (MJD 55113.165, with an
exposure of 3.1 ksec, see Tab.~\ref{table2} for details).

\begin{figure}
  \includegraphics[width=\columnwidth]{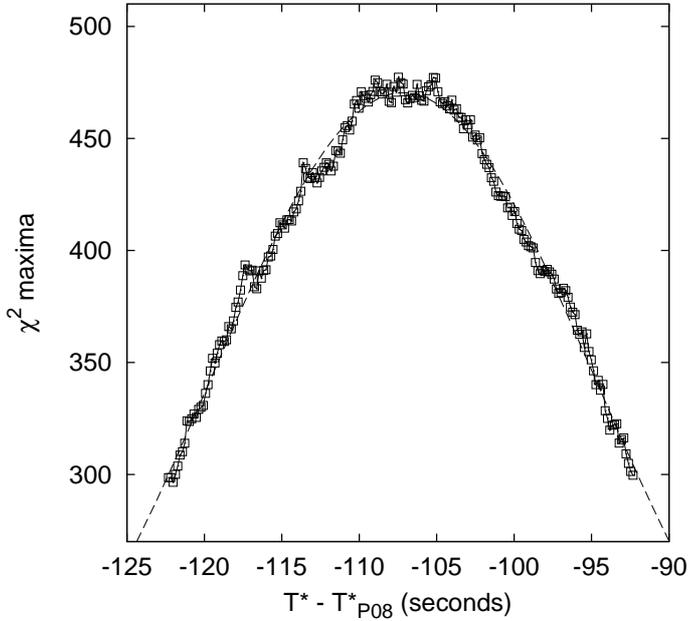}
  \caption{Maximum of the $\chi^2$ obtained in an epoch folding search
    around the expected spin period as a function of the \Tstar
    adopted to correct the time series for the delays induced by the
    orbital motion. In this figure we show the result of the final
    search performed on all the three observations for which the
    pulsation was detected after having corrected for the orbital
    motion using the value \Tstar as estimated from Fig. \ref{fig:f1}.
    A \Tstar step of 0.15 s was adopted and 200 \Tstar were tried.
    The best-fit model, constant + Gaussian, is also shown (dashed
    curve).}
  \label{fig:f3}
\end{figure}

We applied the procedure described above considering all the three
datasets in order to further improve our new $\Tstar_{09}$ measure.
We adopted a finer step in \Tstar of 0.15 s, covering an interval of
30 s around the new value of \Tstar.
Again, we fitted the $\chi^2$ maxima vs. \Tstar curve with a model
consisting of a constant plus a Gaussian, as it is shown in
Fig. \ref{fig:f3}. In this way we were able to obtain a precise
measure of \Tstar;
the final value for \Tstar is reported in Table \ref{table1}.
The evaluation of the uncertainty on \Tstar is discussed in the next
section.

\subsection{Error estimates on \Tstar using Monte Carlo simulations}
The folding search technique described above and adopted to obtain our
refined measure of \Tstar does not provide a straightforward
determination of the uncertainty on this parameter.
To overcome this problem we performed a Monte-Carlo simulation. We
generated 100 datasets with the same exposure, count rate, pulse
modulation and orbital modulation observed in the real data.
For each of these dataset we applied the same procedure as described
in the previous section to obtain a measure of \Tstar.
%
%
The confidence level at 68\% (1$\sigma$) for the \Tstar parameter
corresponds to 1.05 s, which is about 30 times the corresponding
confidence interval obtained by a fit with a gaussian of the $\chi^2$
maxima vs \Tstar curve. We therefore adopt 1.05 s as our best estimate
of the 1~$\sigma$ uncertainty on the \Tstar measure.


\subsection{Timing analysis}
We barycentered our data of \xtejd using our refined orbital solution,
and performed an epoch folding search to determine a mean spin
frequency for the 2009 outburst. From the best-fit of the $\chi^2$
curve we obtained a value of 435.31799237(19) Hz.
We evaluated the frequency uncertainty using the Monte-Carlo simulated
data described above.
We found that the error determined in this way was an order of
magnitude bigger with respect to the error determined fitting with a
gaussian the centroid position of the epoch folding search curve.

\begin{figure}
  \includegraphics[width=\columnwidth]{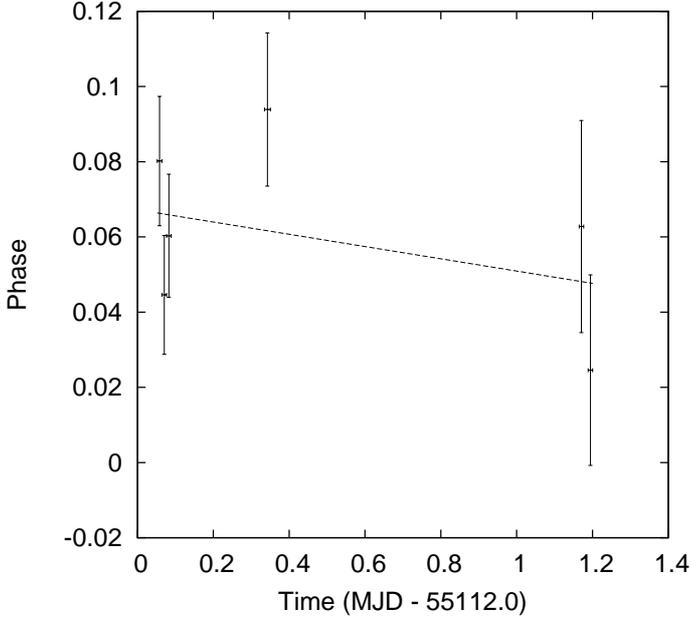}
  \caption{Pulse phase delays of the fundamental for the three
    observations in which the pulsation is detected. Each phase point
    is obtained folding on $\sim$ 1000-s long time intervals and using
    a spin frequency $\nu = 435.31799237$, which is the value obtained
    with the epoch folding search technique. The best-fit constant
    spin frequency model is also shown (dashed line).}
  \label{fig:f4}
\end{figure}

\begin{figure}
  \includegraphics[width=\columnwidth]{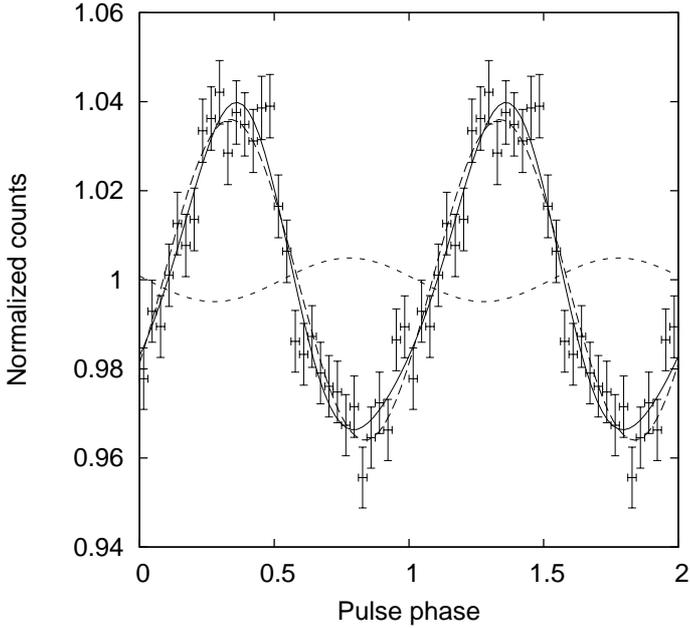}
  \caption{Folded pulse profile of the three datasets in which the
    pulsation was detected. The profile is reported twice for
    clarity.}
  \label{fig:f5}
\end{figure}

Adopting this mean spin frequency value, we epoch folded the three
observations during which pulsations could be detected over 1000 s
long intervals, considering 16 bins to sample the signal (see Table
\ref{table2} for details).
%
%
In this way we obtained 7 folded pulse profiles.
\begin{table}
  \begin{minipage}[t]{\columnwidth}
    \caption{Orbital and Spin Parameters for \xtejd.}
    \label{table1}
    \centering
    \renewcommand{\footnoterule}{}  
    \begin{tabular}{lcl}
      \hline \hline
      Parameter &  Value \\
      \hline
      RA (J2000)  &  $17^{\rm h} 51^{\rm m} 13^{\rm s}\!\!.49(5)$ & $^a$\\
      Dec (J2000) &  $-30^\circ 37{\rm '} 23{\rm ''}\!\!.4(6)$ & $^a$\\
      Projected semi-major axis $a_x \sin i$ (lt-ms)
      &  10.125(5) & $^b$ \\
      Ascending node passage, \Tstar (MJD) at & \\
      \hfill 2002 outburst & 52368.0129023(4) & $^b$ \\
      \hfill 2009 outburst & 55111.000647(12) \\ 
      Orbital period, $P_{orb}$ (s) &  2545.342(2) &  $^b$ \\
      Eccentricity, $\ecc$ & $<$ \np{1.3}{-3} & $^b$ \\

      Reference epoch, T$_0$ (MJD) & 55112.0 \\
      Mean spin frequency, $\nu_0$ (Hz) & 435.31799256(23) \\
      Secular spin frequency derivative, & & \\
      \hfill  $\dot{\nu}_{sec}$ (Hz s$^{-1}$) & \np{-0.55(12)}{-14} & \\
      \hline
    \end{tabular}
    \tablebib{ $^a$\, \citet{markwardt_02}; $^b$\, \citet{papitto_08}}
    \tablefoot{Errors are at $1\sigma$ confidence level, upper limits
      are given at 95\% confidence level. Errors on source position
      are given at 90\% confidence level. Times are referred to the
      barycentre of the Solar System (TDB).}
  \end{minipage}
\end{table}
We performed an harmonic decomposition of each pulse profile.
The fundamental and the first overtone were significantly
detected.
The fundamental was significantly detected on 6 folded pulse profiles
while the first overtone only in one profile.
We fitted the pulse phase delays with a constant plus a linear term,
representing a constant (mean) spin frequency model. 
From the fit we obtained a mean spin frequency of 435.31799256(22),
with a final $\chi^2/dof$ of 1.55(4 dof). This value is in perfect
agreement with the value obtained with the epoch folding search.
Note that the uncertainty on the pulse frequency obtained from the
fitting of the pulse phase delays fit is nearly equal to the
uncertainty on the pulse frequency obtained with a folding search
estimated with the Monte Carlo simulations, so that both procedures
give consistent results.
The pulse phase delays and the best fit line are shown in Figure
\ref{fig:f4}. The pulse profile obtained by folding all the data is
shown in Figure \ref{fig:f5}.

To correctly determine the fractional amplitude, an estimate of the
background and the contribution from sources contaminating the field
of view is mandatory.
In this case, the major contribution to the background is due to the
emission of the AMP \newigrj and of the galactic ridge.
While the roughly constant contribution from the galactic ridge is of
$\sim 7.5$ cts s$^{-1}$ in the considered energy band
\citep{papitto_10}, the \newigrj contribution can only be estimated
extrapolating its flux decay trend just before the \xtejd outburst
onset.
We fitted the 2--18 keV \newigrj X-ray light curve with a linear model
in the time interval from 55108.92 to 55111.46 MJD, as it is possible
to see in Figure \ref{fig:flux}. In Table \ref{table2} we report the
obtained fractional amplitudes for each of the three
observation. These values are corrected with respect to the
instrumental background as well as the estimated contribution from
\newigrj and the galactic ridge, but are still strongly dependent from
the model adopted to describe the \newigrj X-ray light curve.

%
%
%
\changes{The uncertainty on the position of the source quoted by
  \citetalias{markwardt_02} is of 0.6$''$\citep[$90 \%$ confidence
  level, see][]{Aldcroft_00}, while 0.37$''$ is the confidence
  interval corresponding to 1 $\sigma$\footnote{See
    \url{http://cxc.harvard.edu/cal/ASPECT/celmon/}}}.
\changes{Because of this,} a systematic uncertainty arises on the spin
frequency obtained by fitting the pulse phase delays.
These systematic uncertainties are $\sigma_{\nu \;{\rm sys}}$ =
\np{4.8}{-7} $\nu_{3} \; \sigma{''} \; (1 + \sin^2 \beta)^{1/2}$ Hz
\citep{burderi_07}, where $\nu_{3}$ is the spin frequency in units of
1000 Hz, $\sigma{''}$ is the positional error circle in units of
arcsec, and $\beta$ refers to the ecliptic latitude of the source (for
\xtejd $\lambda =268.097281^{\circ}$ and $\beta =-7.198364^{\circ}$).
With these values we have $\sigma_{\nu \;{\rm sys}}$ = \np{7.8}{-8}
Hz.

Combining in quadrature this systematic error with the statistical
error of \np{2.2}{-7} Hz we find a final error on the spin frequency
of $\sigma_{\nu}$ = \np{2.3}{-7} Hz.
Thus the value of the average spin frequency during the 2009 outburst
is $\overline{\nu}_{09} = 435.31799256(23)$ Hz.
\begin{table}
  \begin{minipage}[t]{\columnwidth}
    \caption{Analysed Obs Id exposures and fractional amplitudes.}
    \label{table2}
    \centering
    \renewcommand{\footnoterule}{}  
    \begin{tabular}{lccc}
      \hline \hline
      Obs ID &  Start time & Exposure & fractional amplitude\\
      &  (MJD) & (ks) & (\%) \\
      \hline
      94041-01-04-08 & 55112.052 & 3.2 & 7.9 $\pm$ 1.0 \\
      94041-01-04-04 & 55112.335 & 1.3 & 7.6 $\pm$ 2.2 \\
      94042-01-02-00 & 55113.165 & 3.1 & 7.1 $\pm$ 2.0 \\
      \hline
    \end{tabular}
    \tablefoot{For each Obs Id in which pulsations at the frequency of
      \xtejd have been detected in this work, the start time, the
      exposure and the fractional amplitude (corrected for the
      contribution of \newigrj, galactic ridge, and instrumental
      background) are reported.}
  \end{minipage}
\end{table}

\section{Discussion}  
We obtained a refined orbital solution and a precise estimate of the
spin frequency of the AMP \xtejd from a timing analysis of the RXTE
data during its 2009 outburst.

\subsection{Orbital period evolution}
Although a measure of \Porbt is impossible with only two measurements
of \Tstar, we can derive an upper and lower limit to the \Porbt using
the full orbital solution given by \pap for the 2002 outburst and our
measure of \Tstar for the 2009 outburst.
Using eq.~1 given in \citet{burderi_09}, we can solve that expression
to derive \Porbt
\begin{equation}
  \label{eq:p_orb_dot}
  \Porbt = \frac{2}{N\Porb} \left(\frac{\Delta\Tstar(N)}{N} - \Delta\Porb \right),
\end{equation}
where \Porb is the orbital period measured by \pap, N (= 93109) is the
integer number of orbital cycles between the two \Tstar,
$\Delta\Tstar(N)$ the difference between the measured \Tstar at the
N-th orbital cycle and its expected value, that is $\Delta\Tstar(N) =
\Tstar_{09} - (\Tstar_{P08} + N \times \Porb)$.
%
We can assume that the correction term in Eq. \ref{eq:p_orb_dot},
$\Delta\Porb$, is at most the confidence interval for \Porb given by
\pap. 
Considering the maximum and minimum values of \Porb within its
confidence interval, we obtain that \Porbt lays in the interval from
\np{-2.7}{-11} to \np{+ 0.7}{-11} s s$^{-1}$, at $1\sigma$ confidence
level.

\subsection{Spin frequency secular evolution}
In the following we will derive the secular spin frequency derivative
comparing our measurement of the spin frequency for the 2009 outburst
with the spin frequency measured by \pap analysing the \xtejd 2002
outburst.
Moreover, we will consider possible effects on the spin frequency of
the two outbursts occurred in the intervening seven years.
To compute the effect on the spin frequency of the two weak outbursts
between the 2002 and the 2009 outbursts, we assume that, during each
outburst, the neutron star (NS) is accreting angular momentum $L$ at a
rate
\begin{equation}
  dL/dt = \Mdot\sqrt{GMR_{\rm a}},
  \label{eq:torque}  
\end{equation}
where \Mdot is the mass accretion rate, $G$ is the gravitational
constant, $M$ is the NS mass, $R_{\rm a}$ is the radius at which the
accreting matter (orbiting with Keplerian speed in an accretion disc)
is quickly removed from the disc by the interaction with the NS
magnetic field.
We assume the working hypothesis that the magnetospheric radius
$R_{\rm a}$ can be expressed as (see \citealt{Rappaport_04} and
references therein)
\begin{equation}
  \label{eq:r_alfven}
  R_{\rm a} \propto \Mdot^{-\frac{2}{7}},   
\end{equation}
where $\Mdot$ is the mass accretion rate, and that the mass accretion
rate is proportional to the X-ray flux $F_{\rm X}$.
\changes{As reported by \citet{markwardt_02}, the light-curve of
  \xtejd during the 2002 outburst showed an exponential decay followed
  by a sharp break after which the flux quickly dropped below
  detectability. Therefore we assumed to model the X-ray flux $F_{\rm
    X}$ of each outburst showed by the source with the function}
\begin{equation}
  \label{eq:flux}
  F_{\rm X} = 
  \begin{cases}
    F_{\rm X \; i} \exp{(-t/\tau)}& \text{if $0 \le t \le \bar{t}$}\\
    0& \text{if $t < 0$ or $t > \bar{t}$ },
  \end{cases}
\end{equation}
where $F_{\rm X \; i}$ is the X-ray flux at the outburst peak
\changes{, $\bar{t}$ is the time-length of the exponential decay, }
and $\tau$ the decay time. In Table \ref{table3} $F_{\rm X \; i}$ and
$\tau$ are reported for each outburst.
Since the 2002 outburst is the only one for which a measure of the
spin frequency derivative was possible, we will use it as a reference
for the other outbursts.
Using the above equations and hypotheses, we can derive the spin
frequency derivative and, integrating over time, the spin frequency
variation $\Delta \nu$ in an outburst \changes{of time-length
  $\bar{t}$}.
After some algebraic manipulation, we find
\begin{equation}
  \Delta \nu = \dot{\nu}_{\rm i \; 02} \; \tau^* \left[ 1 - \exp{\left\{-\frac{\bar{t}}{\tau^*}\right\}}\right]  \left(\frac{F_{\rm X \; i}}{F_{\rm X \; i \; 02}}\right)^{\frac{6}{7}}\!\!\!\!,
  \label{eq:delta_nu}
\end{equation}
where $\tau^* = 7\tau / 6$, $F_{\rm X \;i \; 02}$ is the X-ray flux at
the beginning of the 2002 outburst and $\dot{\nu}_{\rm i \; 02}$ is
the corresponding spin frequency derivative.
The flux and the spin frequency derivative of the \changes{2002}
outburst are
$F_{\rm X \;i \; 02} = \np{1.34(7)}{-9}$ erg cm$^{-2}$ s$^{-1}$
\changes{(2--10 keV)},
$\dot{\nu}_{\rm i \; 02}$ = \np{0.56(12)}{-12} Hz s$^{-1}$ (see \pap).\\
\changes{For the same outburst \markw obtained a $\tau = 7.1(1)$ days
  and $\bar{t} = 8.5$ days, whose ratio is $\tau/\bar{t} \simeq
  0.84$.}
\changes{Adopting the same model for the 2009 outburst we found $\tau
  = 2.4(2)$ days and $\bar{t}$ in the range 1.2--2.9 days, which
  implies that the ratio $\tau/\bar{t}$ lies in the range 0.83--2.0.}
\begin{table*}
  \begin{minipage}[t]{0.9\textwidth}
    \caption{Spin frequency values for all the observed \xtejd
      outbursts.}
    \label{table3}
    \centering
    \renewcommand{\footnoterule}{} 
    \begin{tabular}{cccccccc}
      \hline
      \hline
      Outburst & $\nu_{\rm i}$ & $\nu_{\rm f}$ & $\overline{\nu}$ & $F_{\rm X \; i}$ & $\tau$ & $\Delta \nu$ & References \\
      & (Hz) & (Hz) & (Hz) & ($10^{-11}$ erg cm$^{-2}$ s$^{-1}$) & (days) & ($10^{-8}$ Hz) & \\
      \hline
      2002 & 435.31799357(9) &  435.31799385(16) & - & 134(7) & 7.1(1) & 28(8) & $^a$ \\
      \hline          
      2005 & - & - & - &  19(7) &  $\simeq$2.4 & 1.5(4) &  $^b$ \\
      \hline
      2007 & - & - & - & 24(5) & $\simeq$2.3 & 1.7(3) &  $^c$  \\
      \hline
      2009 & 435.31799255(23) & - & 435.31799256(23) & 36(14) & 2.4(2) & 0.8(3)--1.7(6) &  $^d$  \\
      \hline
    \end{tabular}
    \tablebib{ $^a$~\citet{markwardt_02}, \pap;\;
      $^b$~\citet{grebenev_atel_05,swank_atel_05}; \;
      $^c$~\citet{falanga_atel_07,markwardt_atel_07};\; $^d$~this
      work, \citet{chevenez_atel_09}}
    \tablefoot{In this table we report, for each outburst, the spin
      frequency at the start of the outburst ($\nu_{\rm i}$), the spin
      frequency at the end of the outburst ($\nu_{\rm f}$), the
      average spin frequency ($\overline{\nu}$), the flux at the peak
      of the outburst in the energy range 2-10 keV ($F_{\rm X \; i}$),
      the flux decay time scale $\tau$ in the hypothesis of an
      exponential decay, and the inferred spin frequency variation
      during the outburst.}
  \end{minipage}
\end{table*}

\subsubsection{Spin-down between 2002 and 2009 outbursts}
Because a significant spin frequency derivative was detected during
the 2002 outburst (\pap), we considered two frequencies, 
at the beginning and at the end of the 2002 outburst, respectively.
The frequency at the beginning of the 2002 outburst, at $T_0$ =
52368.653 MJD, is $\nu_{\rm 02\; i}$ = 435.31799357(9) Hz.  As usual
the error is on the last digit at $1\sigma$ level and was computed by
combining in quadrature the statistical error on the spin frequency
estimate, \np{4}{-8} Hz, with the systematic error induced by the
uncertainty on the source position, \np{7.8}{-8} Hz.
The frequency at the end of the outburst, which occurred about nine
days after the beginning, when the pulsation was detected for the last
time, is $\nu_{\rm f\;02} = 435.31799385(16)$, which was computed
adopting the mean value for the spin frequency derivative given in
\pap, $\dot{\nu}$ = \np{3.7(1.0)}{-13} Hz s$^{-1}$.  
The error is again computed combining in quadrature the statistical
and systematic uncertainties on $\nu_0$ (\np{4}{-8} Hz and
\np{7.8}{-8} Hz, respectively) with the statistical and systematic
uncertainties in the spin frequency derivative (\np{7.8}{-8} Hz and
\np{1.2}{-8} Hz, respectively), where the systematic error on the spin
frequency derivative induced by the uncertainty on the source position
\citep{burderi_07} is $\sigma_{\dot{\nu} \;{\rm sys}}$ = \np{9.6}{-14}
$\nu_{3} \; \sigma^{''} \; (1 + \sin^2 \beta )^{1/2}$ Hz s$^{-1}$ =
\np{1.6}{-14} Hz s$^{-1}$ for \xtejd.
For the 2009 outburst we considered that $\nu_{09} \simeq
\overline{\nu}_{09}$. 
Averaging Eq. \ref{eq:delta_nu} over the \changes{outburst length
  $\bar{t}$,} we obtain an expression for the spin frequency at the
beginning of the 2009 outburst
%
%
\begin{equation}
  \nu_{\rm i \;09} = \overline{\nu}_{09} - \dot{\nu}_{\rm i \; 02} \; \tau^*  \left\{1 - \frac{\tau^*}{\bar{t}} \left[ 1 - \exp{\left\{-\frac{\bar{t}}{\tau^*}\right\}}\right]\right\}  \left(\frac{F_{\rm X \; i\; 09}}{F_{\rm X \; i \; 02}}\right)^{\frac{6}{7}}.
  \label{eq:nudot}
\end{equation}
\changes{The 2009 outburst length lies in the range 1.2--2.9 days,
  where 1.2 days is the time interval in which the pulsation was
  detected, while 2.9 days is the maximum possible length of the
  outburst (see Fig. \ref{fig:flux}).}
\changes{Assuming $F_{\rm X \; i\; 09} = \np{36}{-11}$ erg cm$^{-2}$
  s$^{-1}$, $\tau = 2.4(2)$ days, and $\bar{t} = 1.2$ Equation
  \ref{eq:nudot} gives} $\nu_{\rm i \;09} = \overline{\nu}_{09}$ -
\np{0.8(3)}{-8} Hz = 435.31799255(23) Hz
\changes{, while considering $\bar{t} = 2.9$ gives $\nu_{\rm i \;09} =
  \overline{\nu}_{09}$ - \np{1.7(6)}{-8} Hz = 435.31799254(23) Hz.}
\changes{The difference between the two cases is, for our purpouses,
  irrelevant.}

To compute the secular spin frequency derivative we consider the total
time elapsed from the end of the 2002 outburst to the beginning of the
2009 outburst, namely $\Delta t = 55112.0 - 52377.653$ MJD.  The
secular spin frequency derivative is
\begin{equation}
  \dot{\nu}_{\rm sec} = \frac{(\nu_{\rm i \;09} - \nu_{\rm f \;02})}{\Delta t} = \np{-0.55(12)}{-14}
  \textrm{Hz s}^{-1},
  \label{eq:spindown_02_09}
\end{equation}
where the $1\sigma$ error is computed by adding in quadrature all the
errors. Note that this value is about one order of magnitude higher
than the secular spin-down rate measured for the AMP \saxj
\citep{hartman_09}, and of the same order of magnitude of the secular
spin-down rate measured for the fastest AMP \igrj
\citep[][]{Patruno_10, Hartman_11, papitto_11b}.
Even in the worst case scenario, that is if we computed the secular
spin frequency derivative neglecting the frequency variation of all
the 2002 outburst (in this case the total time elapsed has been
computed as $\Delta t$ = 55112.0 - 52368.653 MJD, namely the total
time elapsed from the beginning of the 2002 outburst to the beginning
of the 2009 outburst), we obtain $\dot{\nu}_{\rm sec} =$
\np{-0.43(10)}{-14} Hz s$^{-1}$, which is still significant and
compatible, within the errors, with the previous value.

\subsubsection{Effect of the spin-up during 2005 and 2007 outbursts}
We now consider the fact that on 28 March 2005 and on 4 April 2007,
two other outbursts occurred. 
In the following we will discuss the effect of a possible spin-up
during these two outburst, similar to the one observed for the 2002
outburst.
It should be noted here that there is a possibility that the 2005
outburst is not associated to \xtejd, since no precise position for
the source of the outburst is available for the 2005 event.
Using Eq. \ref{eq:delta_nu}, we can evaluate an order of magnitude
estimate of the spin frequency variation that may have occurred during
these two outburst.

For the second outburst, which occurred on 28-29 March 2005, the peak
flux was $\np{19(7)}{-11}$ erg cm$^{-2}$ s$^{-1}$ (2-10 keV), and
$\bar{t} \simeq 2.4$ days \citep{grebenev_atel_05,swank_atel_05}.
%

For the third outburst, 4-5 April 2007, the peak flux was
$\np{24(5)}{-11}$ erg cm$^{-2}$ s$^{-1}$ (2-10 keV), and $\bar{t}
\simeq 2.3$ days \citep{falanga_atel_07,markwardt_atel_07}.
With these values, using Eq. \ref{eq:delta_nu} and adopting
$\bar{t}/\tau \simeq 1$, we derive $\Delta \nu_{05}$ = \np{1.5(4)}{-8} Hz
and $\Delta \nu_{07}$ = \np{1.7(3)}{-8} Hz, where the errors were
computed by propagating the uncertainties on the fluxes.
\changes{The above results are obtained in the hypothesis that
  $\bar{t}/\tau \simeq 1$. The result holds even if we consider, as in
  the case of the 2009 outburst, that $\tau = 2 \bar{t}$. In this case
  the frequency changes in the two outbursts are $\Delta \nu_{05}$ =
  \np{1.8(5)}{-8} Hz and $\Delta \nu_{07}$ = \np{2.1(4)}{-8} Hz.}
It is clear that the effect on the value of $\dot{\nu}_{\rm sec}$,
considering only the 2007 outburst or considering both 2005 and 2007,
is negligible.
We have thus demonstrated that, within the error, the secular spin
frequency derivative is independent on the frequency variations caused
by possible spin-up episodes during the weak 2005 and 2007 outbursts.

\subsection{Magnetic field}
Using the derived secular spin frequency derivative, we can estimate
the magnetic field strength equating the rotational energy loss rate
to the rotating magnetic dipole emission. 
It is not straightforward which expression to use to evaluate the
energy radiated by a rotating dipole. 
While the classical formula for a rotating dipole in vacuum is well
known, an equivalent expression in presence of matter is still debated
in literature.
\citet{goldreich_69} demonstrated that NS typical magnetic field
strengths are strong enough to fill the magnetosphere with charged
particles extracted from the surface, with the result that even an
aligned rotator emits energy. 
We can write the amount of energy radiated as \citep{spitkovsky_06}
\begin{equation}
  \dot{E} \simeq -f(\theta)\; \mu^2\; c^{-3}\;\omega^4, 
  \label{eq:gen_dipole}
\end{equation}
where $c$ is the speed of light, $\mu$ is the magnetic dipole moment,
$\omega$ the NS angular spin frequency, $\theta$ the angle between the
rotation and magnetic axes and $f(\theta)$ a dimensionless function
which takes into account the energy dependence from the angle $\theta$
and the effects of the presence of particles in the magnetosphere. 
In vacuum $f(\theta)$ = {\tiny 2/3} $\sin^2(\theta)$, while in case of
matter in the magnetosphere \citet{spitkovsky_06} proposed, on the
basis of MHD simulations, $f(\theta) = 1 + \sin^2(\theta)$ (see
\citealt{contopoulos_07} for more details).  
Equating the irradiated energy to the rotational energy loss rate we
obtain:
\begin{equation}
  \mu = \sqrt{\frac{I\,c^{3} \dot{\omega}}{f(\theta)\omega^3}} 
  \simeq \np{8.27}{26} f^{-\frac{1}{2}}(\theta)\; I^{\frac{1}{2}}_{45}\; 
  \nu^{-\frac{3}{2}}_2 \dot{\nu}^{\frac{1}{2}}_{-15}\;\; {\rm G\; cm}^{3}, 
  \label{eq:mu2}
\end{equation}
where $I_{45}$ is the NS moment of inertia in units of $10^{45}$ g
cm$^2$, $\nu_2$ the spin frequency in units of 100 Hz and
$\dot{\nu}_{-15}$ the spin frequency derivative in units of $10^{-15}$
Hz s$^{-1}$. 
Using our estimates of the spin frequency and its secular derivative
reported in Table \ref{table1}, we obtain a value for the magnetic
dipole strength of $\mu$ = \np{2.14(23)}{26}
$f^{-\frac{1}{2}}(\theta)\; I^{1/2}_{45} $ G cm$^{3}$.  

\changes{Assuming a 1.4\Msun NS mass and adopting the FPS
  \citep[see][]{friedman_81, pandharipande_89}} equation of state, we
obtain a radius of $R_{\rm NS}$ = \np{1.14}{6} cm and a moment of
inertia $I$ = \np{1.29}{45} gr cm$^2$.
\changes{Under the assumptions above} we can give an estimate of the
magnetic field strength at the magnetic caps $B_{\rm PC}$ from the
relation that gives the dipole magnetic field strength at the NS
surface as a function of the magnetic dipole moment and the angle
$\alpha$ between the position on the surface and the magnetic dipole
axis ($\alpha = 0$ on the magnetic cap) $B = (\mu / R_{\rm
  NS}^3)\sqrt{1 + 3 \cos^2\!\alpha}$.
We find $B_{\rm PC} = \np{3.3(4)}{8} f^{-\frac{1}{2}}(\theta)\; {\rm
  G}$.  
Adopting $f(\theta) = 2/3$ (in line with what is assumed in deriving
the magnetic field of radio pulsars), we find $B_{\rm PC} =
\np{4.0(4)}{8} \; {\rm G}$, which is quite reasonable for this kind of
source.

\citet{Hartman_11} noted that the AMPs for which the secular spin-down
was measured are suitable to be detected as $\gamma$-ray millisecond
pulsars by the \fermi Large Area Telescope, since several millisecond
radio pulsars with similar characteristics were detected
\citep[see][]{Abdo_09}.
The spin-down power, defined as $\dot{\rm E} = 4 \pi^2 I \omega
\dot{\omega}$, for this source is $\dot{\rm E} = \np{0.9(2)}{35}
I_{45}$ erg s$^{-1}$.
Following \citet{Abdo_09}, the upper limit on the $\gamma$-ray flux is
$\eta \dot{\rm E}/d^2 \leq \np{2.4}{33} I_{45}$ erg s$^{-1}$
kpc$^{-2}$, where $\eta$ is the $\gamma$-ray efficiency
\citep[see][for $\eta$ definition]{Abdo_09} and $d$ is the \xtejd
distance lower limit, estimated by \pap to be 6.3 kpc.
The observed values for $\eta$ lays in the range 6 -- 100 \%
\citep[][]{Abdo_09}.
The chance to detect \xtejd as $\gamma$-ray is then unlikely, although
possible if the $\gamma$-ray efficiency is high.

For completeness, NS magneto-dipole radiation is not the only process
to invoke for NS angular momentum loss. NS mass distribution can
deviate from a perfectly spherical distribution for several reasons
\citep[see, e.g.][]{Bildsten_98a}, introducing a neutron star's mass
quadrupole moment which permits emission of gravitational waves (GW)
at a frequency double of the NS spin frequency.
Following \citet{Hartman_11} and \citet{papitto_11b}, it is possible
to give an upper limit to the average neutron star's mass quadrupole
moment Q, under the hypothesis that the spin down is due only to GW
emission.
Using the expression for the net torque due to a mass quadrupole
moment given in \citet{Thorne_80} and adopting the value of the spin
frequency and its derivative obtained in this work, we can derive an
upper limit to the quadrupole ellipticity \citep[see,
e.g.][]{Ferrari_10} $Q/I \leq \np{4.6}{-9}\; I^{-1/2}_{45}$ (3$\sigma$
confidence level), in line with the values obtained for the sources
\saxj \citep{Hartman_08} and \igrj \citep{Hartman_11,papitto_11b}. It
it also in agreement with the upper limits obtained for the
millisecond radio pulsars.
With these values of ellipticity and source distance there is no
chance it can be detected by the current GW detectors
\citep[e.g.][]{Abbott_10}, since their sensitivity is orders of
magnitude worse.

\begin{acknowledgements} 
  We thank A. Possenti for some fruitful discussions and the unknown 
  referee for useful suggestions.
  This work is supported by the Italian Space Agency, ASI-INAF
  I/088/06/0 contract for High Energy Astrophysics.
\end{acknowledgements}

\bibliography{ms}

\begin{thebibliography}{37}
\expandafter\ifx\csname natexlab\endcsname\relax\def\natexlab#1{#1}\fi

\bibitem[{{Abbott} {et~al.}(2010){Abbott}, {Abbott}, {Acernese}, {Adhikari},
  {Ajith}, {Allen}, {Allen}, {Alshourbagy}, {Amin}, {Anderson}, \&
  et~al.}]{Abbott_10}
{Abbott}, B.~P., {Abbott}, R., {Acernese}, F., {et~al.} 2010, \apj, 713, 671

\bibitem[{{Abdo} {et~al.}(2009){Abdo}, {Ackermann}, {Ajello}, {Atwood},
  {Axelsson}, {Baldini}, {Ballet}, {Barbiellini}, {Baring}, {Bastieri},
  {Baughman}, {Bechtol}, {Bellazzini}, {Berenji}, {Bignami}, {Blandford},
  {Bloom}, {Bonamente}, {Borgland}, {Bregeon}, {Brez}, {Brigida}, {Bruel},
  {Burnett}, {Caliandro}, {Cameron}, {Camilo}, {Caraveo}, {Carlson},
  {Casandjian}, {Cecchi}, {{\c C}elik}, {Charles}, {Chekhtman}, {Cheung},
  {Chiang}, {Ciprini}, {Claus}, {Cognard}, {Cohen-Tanugi}, {Cominsky},
  {Conrad}, {Corbet}, {Cutini}, {Dermer}, {Desvignes}, {de Angelis}, {de Luca},
  {de Palma}, {Digel}, {Dormody}, {do Couto e Silva}, {Drell}, {Dubois},
  {Dumora}, {Edmonds}, {Farnier}, {Favuzzi}, {Fegan}, {Focke}, {Frailis},
  {Freire}, {Fukazawa}, {Funk}, {Fusco}, {Gargano}, {Gasparrini}, {Gehrels},
  {Germani}, {Giebels}, {Giglietto}, {Giordano}, {Glanzman}, {Godfrey},
  {Grenier}, {Grondin}, {Grove}, {Guillemot}, {Guiriec}, {Hanabata}, {Harding},
  {Hayashida}, {Hays}, {Hobbs}, {Hughes}, {J{\'o}hannesson}, {Johnson},
  {Johnson}, {Johnson}, {Johnson}, {Johnston}, {Kamae}, {Katagiri}, {Kataoka},
  {Kawai}, {Kerr}, {Kn{\"o}dlseder}, {Kocian}, {Kramer}, {Kuss}, {Lande},
  {Latronico}, {Lemoine-Goumard}, {Longo}, {Loparco}, {Lott}, {Lovellette},
  {Lubrano}, {Madejski}, {Makeev}, {Manchester}, {Marelli}, {Mazziotta},
  {McConville}, {McEnery}, {McLaughlin}, {Meurer}, {Michelson}, {Mitthumsiri},
  {Mizuno}, {Moiseev}, {Monte}, {Monzani}, {Morselli}, {Moskalenko}, {Murgia},
  {Nolan}, {Norris}, {Nuss}, {Ohsugi}, {Omodei}, {Orlando}, {Ormes}, {Paneque},
  {Panetta}, {Parent}, {Pelassa}, {Pepe}, {Pesce-Rollins}, {Piron}, {Porter},
  {Rain{\`o}}, {Rando}, {Ransom}, {Ray}, {Razzano}, {Rea}, {Reimer}, {Reimer},
  {Reposeur}, {Ritz}, {Rochester}, {Rodriguez}, {Romani}, {Roth}, {Ryde},
  {Sadrozinski}, {Sanchez}, {Sander}, {Saz Parkinson}, {Scargle}, {Schalk},
  {Sgr{\`o}}, {Siskind}, {Smith}, {Smith}, {Spandre}, {Spinelli}, {Stappers},
  {Starck}, {Striani}, {Strickman}, {Suson}, {Tajima}, {Takahashi}, {Tanaka},
  {Thayer}, {Thayer}, {Theureau}, {Thompson}, {Thorsett}, {Tibaldo}, {Torres},
  {Tosti}, {Tramacere}, {Uchiyama}, {Usher}, {Van Etten}, {Vasileiou},
  {Venter}, {Vilchez}, {Vitale}, {Waite}, {Wallace}, {Wang}, {Watters}, {Webb},
  {Weltevrede}, {Winer}, {Wood}, {Ylinen}, \& {Ziegler}}]{Abdo_09}
{Abdo}, A.~A., {Ackermann}, M., {Ajello}, M., {et~al.} 2009, Science, 325, 848

\bibitem[{{Aldcroft} {et~al.}(2000){Aldcroft}, {Karovska},
  {Cresitello-Dittmar}, {Cameron}, \& {Markevitch}}]{Aldcroft_00}
{Aldcroft}, T.~L., {Karovska}, M., {Cresitello-Dittmar}, M.~L., {Cameron},
  R.~A., \& {Markevitch}, M.~L. 2000, in Society of Photo-Optical
  Instrumentation Engineers (SPIE) Conference Series, Vol. 4012, Society of
  Photo-Optical Instrumentation Engineers (SPIE) Conference Series, ed.
  {J.~E.~Truemper \& B.~Aschenbach}, 650--657

\bibitem[{{Altamirano} {et~al.}(2010){Altamirano}, {Patruno}, {Heinke},
  {Linares}, {Markwardt}, \& {Strohmayer}}]{altamirano_atel_10}
{Altamirano}, D., {Patruno}, A., {Heinke}, C.~O., {et~al.} 2010, The
  Astronomer's Telegram, 2500, 1

\bibitem[{{Bildsten}(1998)}]{Bildsten_98a}
{Bildsten}, L. 1998, \apjl, 501, L89+

\bibitem[{{Burderi} {et~al.}(2007){Burderi}, {Di Salvo}, {Lavagetto}, {Menna},
  {Papitto}, {Riggio}, {Iaria}, {D'Antona}, {Robba}, \& {Stella}}]{burderi_07}
{Burderi}, L., {Di Salvo}, T., {Lavagetto}, G., {et~al.} 2007, \apj, 657, 961

\bibitem[{{Burderi} {et~al.}(2009){Burderi}, {Riggio}, {di Salvo}, {Papitto},
  {Menna}, {D'A{\`i}}, \& {Iaria}}]{burderi_09}
{Burderi}, L., {Riggio}, A., {di Salvo}, T., {et~al.} 2009, \aap, 496, L17

\bibitem[{{Chenevez} {et~al.}(2009){Chenevez}, {Kuulkers}, {Beckmann}, {Bird},
  {Brandt}, {Domingo}, {Ebisawa}, {Jonker}, {Kretschmar}, {Markwardt},
  {Oosterbroek}, {Paizis}, {Risquez}, {Sanchez-Fernandez}, {Shaw}, \&
  {Wijnands}}]{chevenez_atel_09}
{Chenevez}, J., {Kuulkers}, E., {Beckmann}, V., {et~al.} 2009, The Astronomer's
  Telegram, 2235, 1

\bibitem[{{Contopoulos}(2007)}]{contopoulos_07}
{Contopoulos}, I. 2007, in Proceedings of the 363. WE-Heraeus Seminar on
  Neutron Stars and Pulsars 40 years after the discovery. Edited by W. Becker
  and H. H. Huang. MPE-Report 291. ISSN 0178-0719. Published by the Max Planck
  Institut f{\"u}r extraterrestrische Physik, Garching bei M{\"u}nchen,
  Germany, 2007., p.134, ed. {W.~Becker \& H.~H.~Huang}, 134--+

\bibitem[{{di Salvo} {et~al.}(2008){di Salvo}, {Burderi}, {Riggio}, {Papitto},
  \& {Menna}}]{disalvo_08}
{di Salvo}, T., {Burderi}, L., {Riggio}, A., {Papitto}, A., \& {Menna}, M.~T.
  2008, \mnras, 389, 1851

\bibitem[{{Falanga} {et~al.}(2007){Falanga}, {Soldi}, {Shaw}, {Goldwurm},
  {Belanger}, {Porquet}, {Melia}, {Terrier}, \&
  {Yusef-Zadeh}}]{falanga_atel_07}
{Falanga}, M., {Soldi}, S., {Shaw}, S., {et~al.} 2007, The Astronomer's
  Telegram, 1046, 1

\bibitem[{{Ferrari}(2010)}]{Ferrari_10}
{Ferrari}, V. 2010, Classical and Quantum Gravity, 27, 194006

\bibitem[{{Friedman} \& {Pandharipande}(1981)}]{friedman_81}
{Friedman}, B. \& {Pandharipande}, V.~R. 1981, Nuclear Physics A, 361, 502

\bibitem[{{Galloway} {et~al.}(2008){Galloway}, {Hartman}, {Chakrabarty},
  {Morgan}, \& {Swank}}]{galloway_atel_08}
{Galloway}, D.~K., {Hartman}, J.~M., {Chakrabarty}, D., {Morgan}, E.~H., \&
  {Swank}, J.~H. 2008, The Astronomer's Telegram, 1786, 1

\bibitem[{{Galloway} {et~al.}(2005){Galloway}, {Markwardt}, {Morgan},
  {Chakrabarty}, \& {Strohmayer}}]{galloway_05}
{Galloway}, D.~K., {Markwardt}, C.~B., {Morgan}, E.~H., {Chakrabarty}, D., \&
  {Strohmayer}, T.~E. 2005, \apjl, 622, L45

\bibitem[{Goldreich \& Julian(1969)}]{goldreich_69}
Goldreich, P. \& Julian, W.~H. 1969, \apj, 157, 869

\bibitem[{{Grebenev} {et~al.}(2005){Grebenev}, {Molkov}, \&
  {Sunyaev}}]{grebenev_atel_05}
{Grebenev}, S.~A., {Molkov}, S.~V., \& {Sunyaev}, R.~A. 2005, The Astronomer's
  Telegram, 446, 1

\bibitem[{{Hartman} {et~al.}(2011){Hartman}, {Galloway}, \&
  {Chakrabarty}}]{Hartman_11}
{Hartman}, J.~M., {Galloway}, D.~K., \& {Chakrabarty}, D. 2011, \apj, 726, 26

\bibitem[{{Hartman} {et~al.}(2008){Hartman}, {Patruno}, {Chakrabarty},
  {Kaplan}, {Markwardt}, {Morgan}, {Ray}, {van der Klis}, \&
  {Wijnands}}]{Hartman_08}
{Hartman}, J.~M., {Patruno}, A., {Chakrabarty}, D., {et~al.} 2008, \apj, 675,
  1468

\bibitem[{{Hartman} {et~al.}(2009){Hartman}, {Patruno}, {Chakrabarty},
  {Markwardt}, {Morgan}, {van der Klis}, \& {Wijnands}}]{hartman_09}
{Hartman}, J.~M., {Patruno}, A., {Chakrabarty}, D., {et~al.} 2009, \apj, 702,
  1673

\bibitem[{{Jahoda} {et~al.}(2006){Jahoda}, {Markwardt}, {Radeva}, {Rots},
  {Stark}, {Swank}, {Strohmayer}, \& {Zhang}}]{Jahoda_06}
{Jahoda}, K., {Markwardt}, C.~B., {Radeva}, Y., {et~al.} 2006, \apjs, 163, 401

\bibitem[{{Kirsch} {et~al.}(2004){Kirsch}, {Mukerjee}, {Breitfellner},
  {Djavidnia}, {Freyberg}, {Kendziorra}, \& {Smith}}]{kirsch_04}
{Kirsch}, M.~G.~F., {Mukerjee}, K., {Breitfellner}, M.~G., {et~al.} 2004, \aap,
  423, L9

\bibitem[{{Markwardt} {et~al.}(2009){Markwardt}, {Altamirano}, {Strohmayer}, \&
  {Swank}}]{markwardt_atel_09b}
{Markwardt}, C.~B., {Altamirano}, D., {Strohmayer}, T.~E., \& {Swank}, J.~H.
  2009, The Astronomer's Telegram, 2237, 1

\bibitem[{{Markwardt} {et~al.}(2007){Markwardt}, {Pereira}, \&
  {Swank}}]{markwardt_atel_07b}
{Markwardt}, C.~B., {Pereira}, D., \& {Swank}, J.~H. 2007, The Astronomer's
  Telegram, 1051, 1

\bibitem[{{Markwardt} \& {Swank}(2007)}]{markwardt_atel_07}
{Markwardt}, C.~B. \& {Swank}, J.~H. 2007, The Astronomer's Telegram, 1045, 1

\bibitem[{{Markwardt} {et~al.}(2002){Markwardt}, {Swank}, {Strohmayer}, {in 't
  Zand}, \& {Marshall}}]{markwardt_02}
{Markwardt}, C.~B., {Swank}, J.~H., {Strohmayer}, T.~E., {in 't Zand},
  J.~J.~M., \& {Marshall}, F.~E. 2002, \apjl, 575, L21

\bibitem[{{Pandharipande} \& {Ravenhall}(1989)}]{pandharipande_89}
{Pandharipande}, V.~R. \& {Ravenhall}, D.~G. 1989, in NATO ASIB Proc. 205:
  Nuclear Matter and Heavy Ion Collisions, ed. {M.~Soyeur, H.~Flocard,
  B.~Tamain, \& M.~Porneuf}, 103--+

\bibitem[{{Papitto} {et~al.}(2008){Papitto}, {Menna}, {Burderi}, {di Salvo}, \&
  {Riggio}}]{papitto_08}
{Papitto}, A., {Menna}, M.~T., {Burderi}, L., {di Salvo}, T., \& {Riggio}, A.
  2008, \mnras, 383, 411

\bibitem[{{Papitto} {et~al.}(2011){Papitto}, {Riggio}, {Burderi}, {di Salvo},
  {D'A{\'{\i}}}, \& {Iaria}}]{papitto_11b}
{Papitto}, A., {Riggio}, A., {Burderi}, L., {et~al.} 2011, \aap, 528, A55+

\bibitem[{{Papitto} {et~al.}(2010){Papitto}, {Riggio}, {di Salvo}, {Burderi},
  {D'A{\`i}}, {Iaria}, {Bozzo}, \& {Menna}}]{papitto_10}
{Papitto}, A., {Riggio}, A., {di Salvo}, T., {et~al.} 2010, \mnras, 407, 2575

\bibitem[{{Patruno}(2010)}]{Patruno_10}
{Patruno}, A. 2010, \apj, 722, 909

\bibitem[{{Patruno} {et~al.}(2010){Patruno}, {Altamirano}, \&
  {Messenger}}]{Patruno_10b}
{Patruno}, A., {Altamirano}, D., \& {Messenger}, C. 2010, \mnras, 403, 1426

\bibitem[{{Rappaport} {et~al.}(2004){Rappaport}, {Fregeau}, \&
  {Spruit}}]{Rappaport_04}
{Rappaport}, S.~A., {Fregeau}, J.~M., \& {Spruit}, H. 2004, \apj, 606, 436

\bibitem[{{Riggio} {et~al.}(2011){Riggio}, {Papitto}, {Burderi}, {di Salvo},
  {Bachetti}, {Iaria}, {D'A{\`i}}, \& {Menna}}]{riggio_11}
{Riggio}, A., {Papitto}, A., {Burderi}, L., {et~al.} 2011, \aap, 526, A95+

\bibitem[{Spitkovsky(2006)}]{spitkovsky_06}
Spitkovsky, A. 2006, \apjl, 648,

\bibitem[{{Swank} {et~al.}(2005){Swank}, {Markwardt}, \&
  {Smith}}]{swank_atel_05}
{Swank}, J.~H., {Markwardt}, C.~B., \& {Smith}, E.~A. 2005, The Astronomer's
  Telegram, 449, 1

\bibitem[{{Thorne}(1980)}]{Thorne_80}
{Thorne}, K.~S. 1980, Reviews of Modern Physics, 52, 299

\end{thebibliography}

\end{document}